\documentstyle[11pt,newpasp,twoside]{article}
\def\gsim{\raise2.90pt\hbox{$\scriptstyle>$} \hspace{-6.4pt}
\lower.5pt\hbox{$\scriptscriptstyle
\sim$}\; }
\def\lsim{\raise2.90pt\hbox{$\scriptstyle<$} \hspace{-6pt}\lower.5pt\hbox{$\scriptscriptstyle\sim$}\; }
\def\be{\begin{equation}}
\def\ee{\end{equation}}

\def\BV{{\bf V}}
\def\etal{{\it et al.}\ }

\def\E{{\bf E}}

\def\beq{\begin{eqnarray}}
\def\eeq{\end{eqnarray}}
\def\nn{\nonumber}

\def\A{{\bf A}}
\def\lra#1{\left\langle #1\right\rangle}
\def\bv{\bf v}
\def\OB{\overline {\bf B}}
\def\bbb{\overline {\bf B}}
\def\bbV{\overline {\bf V}}
\def\bbB{\overline {\bf B}}
\def\bbE{\overline {\bf E}}
\def\bbA{\overline {\bf A}}

\def\cnt{\cdot\nabla\times}

\def\ob{\bar{B}}
\def\ao{\alpha\hbox{-}\Omega}
\def\B{{\bf B}}
\def\A{{\bf A}}
\def\E{{\bf E}}
\def\bfb{\bf b}
\def\bfe{\bf e}
\def\bfj{\bf j}
\def\bfa{\bf a}
\def\bfv{\bf v}
\def\ts{\times}

\def\rb{\rangle}
\def\lb{\langle}
\def\curl{\nabla \times}

\def\edcomment#1{\iffalse\marginpar{\raggedright\sl#1\/}\else\relax\fi}
\reversemarginpar

\begin{document}
\title{Mean Field Dynamo Saturation: Toward Understanding Conflicting
Results}

\author{Eric G. Blackman}
\affil{Department of Physics \& Astronomy, 
University of Rochester, Rochester, NY 91125}

 \author{George B. Field}
\affil{Harvard-Smithsonian Center for Astrophysics,
Cambridge, MA 02138}

\begin{abstract}
Mean field dynamos may explain the origin of 
large scale magnetic fields of galaxies, 
but controversy arises over the extent of dynamo quenching by 
the growing field. Here we explain how apparently conflicting results may  
be mutually consistent, by showing the role of  
magnetic helicity conservation and boundary terms usually neglected.
We estimate the associated magnetic energy flowing out of 
the Galaxy but emphasize that the mechanism of field escape needs 
to be addressed. 
\end{abstract}

\section{Field Growth and Constraining the Turbulent EMF}

Unlike the  turbulent amplification of small scale 
magnetic energy to near equipartition with the the kinetic energy spectrum, 
the mean field dynamo (MFD) 
field generation (c.f. Parker 1979, Kulsrud 1999) 
on scales $>$ turbulent input scale is controversial
(c.f. Field et al. 1999). 
The MFD equation is $\partial_t\bbb=\curl \lb\bfv \ts \bfb \rb+\lambda\nabla^2 \bbb + \curl (\bbV\ts \bbb)$, with the turbulent EMF 
$\lb\bfv\ts\bfb\rb
=\alpha\bbb-\beta\curl\bbb$, 
and pseudo-scalar   $\alpha$ and scalar $\beta$.  
How well MFD growth applies when the when the dynamic magnetic  
backreaction is included depends on the survival of
$\lb\bfv\ts\bfb\rb$.
Blackman \& Field (2000a) used Ohm's law and mean field theory
(e.g.${\bf B}={\bfb}+{\overline {\bf B}}$; \ $\langle\bfb\rangle$ =0)
to constrain the dynamic value of $\lb \bfv\ts \bfb\rb$ analytically.
Deriving  
$\lb\bfv\ts\bfb\rb\cdot{\bbB}
/c = -\eta\lb\bfj\cdot\bfb\rb +\lb\bfe\cdot\bfb\rb$
and then expanding the fluctuating electric field $\bfe$ into its
potentials, gives 
$\lb\bfe\cdot\bfb\rb = -\partial_t\lb\bfa\cdot\bfb\rb/2c+\nabla\cdot
\lb\bfa\ts \bfe -\phi\bfb\rb$.
Thus, for $\lb\bfv\ts\bfb\rb$ not to be resistively 
limited, there must be time variation of 
$\lb \bfa\cdot\bfb\rb$, or non-vanishing boundary terms.  When 
such terms vanish, helical turbulence without mean field gradients 
gives $\alpha \le (b/{\bar B})^2 \alpha_0/R_m$, where
$\alpha_0$ is the kinematic value of the psuedoscalar  
coefficient $\alpha$, and $R_m$ is the magnetic Reynolds number.
There is thus an ambiguity in interpreting all existing
numerical experiments suggesting $\alpha$ quenching (e.g. Cattaneo \& Hughes 1996); the quenching
might not be dynamical, but may be due to boundary conditions.

\section{Magnetic Helicity Escape, Dynamo Action, \& Coronal Activity}
The above result highlights the role of total magnetic helicity 
$H^M = \int_V \A\cdot \B\, d^3x$ (Els\"asser 1956), 
where $V$ is a volume of integration,  
and $\A$ is the vector potential.
That MFD growth involves a magnetic helicity inverse cascade 
was demonstrated by Pouquet \etal (1976).  
The $\alpha$ effect conserves $H^M$
by pumping a positive (negative) amount to scales $>L$ (the outer
turbulent scale) and a negative (positive) amount to
scales $\ll L$.  Brandenburg's (2000) simulations confirm this inverse cascade
and the role of $H^M$ conservation.

A large-scale field can be generated only as fast 
as $H_M$ can be removed or dissipated. Presently, simulations have invoked
boundary conditions for which the growth of large scale field is 
resisitvely limited.  Large $R_m$ systems
must rely on open boundary conditions. 
To see this, note that $H^M$ satisfies 
$
{\partial_t} (\A\cdot\B)+c \nabla\cdot (\E\times \A+A_0\B) =
-2c\E\cdot \B
$ 
where $\E = -{\BV\over c}\times \B $.
Consider two cases.
Case (1): The mean scale = universal scale, or 
the integration is over periodic boundaries.  Then  boundary terms 
vanish, so $\partial_t \lb{\bf A}\cdot {\bf B} \rb=
-2c \lb{\bf E}\cdot {\bf B}\rb=
-2c {\bbE}\cdot {\bbB}-2c \lb{\bfe}\cdot {\bfb}\rb=\eta \lb {\bf J}\cdot{\bf B}\rb$
and
$\partial_t ({\bbA}\cdot \bbB) =-2c {\overline {\bf E}}\cdot {\bbB}; \ \  \partial_t \lb{\bfa}\cdot {\bfb} \rb=-2c \lb {\bfe}\cdot {\bfb}\rb$.
This is the case of section 1. Dynamo action is resistively limited. Case (2): The system (e.g. Galaxy or Sun) mean volume V $<<$  universal volume.
Here we must use the gauge invariant
relative helicity $H^M_R$ inside and outside
of the spherical or disk rotator (Berger \& Field 1984).  
The integral over the the universal volume then satisfies
$\partial_t\int_{U}{\bf A} \cdot {\bf B}d^3x=\partial_t{H}^M_{R,in}+\partial_t{H}^M_{R,out}=-2c\int_U{\bf E}\cdot{\bf B}d^3x\simeq 0$.  The formulae for the $H^M_R$ of the mean and fluctuating quantities inside the rotator 
are $\partial_t {{H}}_{R,in}(\bbB)=
-2c\int_{in}{\bbE\cdot\bbB} d^3x
+2c\int_{S_{in}}({\bbA}_p\times 
{\bbE})\cdot d{\bf S}
$
and 
$
\partial_t{{H}}_{R,in}({\bfb})=
-2c\int_{in}{\lb\bfe\cdot\bfb\rb} d^3x
+2c\int_{S_{in}}\lb{\bf a}_p\times 
{\bfe}\rb\cdot d{\bf S}.
\label{r4aae}
$
In a steady state,  $\partial_tH_{R,in}=0=\partial_tH_{R,out}$.
This and ${\bf E}\cdot{\bf B}\simeq 0$ imply that the above surface
terms above must be equal and opposite.  
Moreover, the surface term balances the $\bbE\cdot \bbB$ 
term in the $\partial_tH^M_{R,in}$ equation.
The boundary term can thus allow for a significant turbulent EMF
because the latter is contained in $\bbE\cdot \bbB$.
Dynamo action unrestricted by 
resisitivity is possible only  in case (2).
This is consistent with Pouquet et al. (1976) and 
Brandenburg (2000).  

If $H_M$ flows through the boundary, then so does magnetic magnetic energy.  
Blackman \& Field (2000b) showed that 
%for field geometries with separable time dependences,  
a typical minimum power leaving the system
when a MFD is operating is given by
 $\dot E^M 
\ge 
{k_{\rm min}\over
8\pi} |\dot H^M|\nn
= {k_{\rm min}\over 6} \left| \lra{\alpha\ob^2}\right|V
$, where $V$ is the system volume.
Dynamos operating 
in the Sun, accretion disks, and the Galaxy would then 
lead to a net escape of magnetic 
energy and small and large scale magnetic helicity. 
Coronal activity from the emergence and
dissipation of helical magnetic flux is thus a prediction
of the MFD in all of these cases, and is observed directly in the Sun 
(c.f. Pevtsov et al. 1999). 
For the Galaxy, \noindent ${\dot E}^M \gsim ({\pi R^2})\alpha \ob^2
\sim10^{40} ({R / 12{\rm kpc}})^2 ({\alpha / 10^5{\rm cm/s}})
({\ob/ 5\ts 10^{-6}{\rm G}})^2 {\rm erg/s}$, in each hemisphere.
Blackman \& Field (2000b) discuss how 
this relation may be 
consistent with coronal energy input rates required by Savage (1995)
and Reynolds et al. (1999).

\section{Open Questions}

An MFD unlimited by resitivivity 
requires the helicity to flow through the boundary
AND that there be some mechanism that enables this flow. Thus there
are two separate issues. Even if the boundary conditions allow it, 
does it actually happen? One may have to include the dynamics of buoyancy 
or winds to fully demonstrate the non-resistive MFD.
Note that turbulent diffusion of the mean
magnetic field (not necessarily the actual field) across the boundary 
is required to maintain a quadrupole field in the Galaxy with a net
flux inside the disk. Similarly, for the Sun, the solar cycle requires
net diffusion through the boundary. The flow of helicity 
would appeal to the same dynamics needed by these constraints.

The analytic and numerical studies that we have seen which 
show catastrophic suppression of the dynamo coefficients, or
resistively limited dynamo action, 
either (1) invoke periodic boundary conditions, and/or (2) 
are 2-D, or (3) do not distinguish between zeroth order isotropic
components of the turbulence and the higher order anistropic
perturbations for a weak mean field (Blackman \& Field 1999). 
This means that there always seems to be an alternative explanation.
and the observed suppression is then ambiguous 
as we have described.  This does not mean that some of the physical concepts
found in the strong suppression results are invalid, but just that they 
may be valid only for the restricted cases considered.  
For example, the observation that the Lagrangian chaos properties of the flow
are changed in the presence of a weak mean field for turbulence in a 
periodic box (e.g. Cattaneo et al. 1996)
needs to be understood in the relation to 
the imposed boundary conditions, and the
shape of the magnetic energy spectrum (e.g. dominated at small or large scales?).  Along these lines, note that the helicity constraint
is global, but also becomes a constraint for any sub-volume
of a periodic box once the system is fully mixed.

\section{References}
%\begin{enumerate}
\noindent
Blackman, E. G. \& Field, G. B. 1999, ApJ {\bf 521} 597.

\noindent
Blackman, E. G. \& Field, G. B. 2000a, ApJ {\bf 534} 984.

\noindent
Blackman, E. G. \& Field, G. B. 2000b, MNRAS, in press,
astro-ph/9912459.

\noindent Berger M.C. \& Field G.B., 1984, JFM, {\bf 147} 133.

\noindent Brandenburg A., 2000, submitted to ApJ, astro-ph/0006186.

\noindent
Cattaneo, F., \& Hughes, D.W. 1996, Phys.\ Rev.\ E.\ {\bf 54}, 4532.

\noindent
Cattaneo, F., \& Hughes, D.W., Kim E.-J, 1996, PRL {\bf 76}, 2057.

\noindent
Els\"asser, W. M. 1956, Rev.\ Mod.\ Phys.\ {\bf 23}, 135.

\noindent
Field, G. 1986, 
%Magnetic Helicity in Astrophysics, 
in {\it
Magnetospheric Phenomena in Astrophysics}, R. Epstein \& W. Feldman, eds.,
AIP Conf. Proc. 144 (Los Alamos: Los Alamos Sci. Lab.), 324.

\noindent
Field, G. B., Blackman, E. G., \& Chou, H. 1999, ApJ {\bf 513}, 638.

\noindent
Frisch, U., Pouquet, A., L\'eorat, J. \& Mazure, A. 1975, JFM {\bf
68}, 769.

%\noindent
%Gruzinov, A. \& Diamond, P. H. 1994, PRL {\bf 72}, 1651.

\noindent Krause F. \& R\"adler K.-H., 1980, {\it 
Mean-field magnetohydrodynamics and dynamo theory}, (New York: Pergamon).

\noindent Kulsrud R., 1999, ARAA, {\bf 37} 37.

\noindent
Parker, E. N., {\it Cosmical Magnetic Fields} (Oxford: Clarendon
Press).

%\noindent
%Parker, E. N. 1955, ApJ {\bf 122}, 293.

\noindent
Pevtsov, A., Canfield, R. C., \& Brown, M. R. eds.\ 1999, {\it Magnetic
Helicity in Space and Laboratory Plasmas}, (American Geophysysical Union).

\noindent
Pouquet, A., Frisch, U., \& Leorat, J. 1976, JFM {\bf 77}, 321.

\noindent
Reynolds R.J., Haffner L.M., Tufte S.L., 1999, ApJ {\bf 525} 21.

\noindent Savage B.D., 1995, in {\it The Physics of the Interstellar Medium
and Intergalactic Medium},  A. Ferrara, C.F. McKee, C. Heiles, \& 
P.R. Shapiro eds. ASP conf ser vol 60. (San Francisco:  PASP) p233.

%\noindent
%Seehafer, N. 1994, Europhys.\ Lett.\ {\bf 27}, 353.

%\noindent
%Shibata, K. 1999, in {\it Magnetic Helicity in Space and
%Laboratory Plasmas}, A. Pevtsov, R. C. Canfield, and M. R. Brown eds.\ (American
%Geophysical Union, 1999).

%\noindent
%Steenbeck, M., Krause, F., \& R\"adler, K. H. 1966, Z. Naturforsch.
%{\bf 21a}, 369.

%\noindent
%Withbroe, G. L. \& Noyes, R. W. 1977, Ann.\ Rev.\ Astron.\
%Astrophys.\ {\bf 15}, 363.
%\end{enumerate}
\end{document}